\catcode`\@=11					



\font\fiverm=cmr5				
\font\fivemi=cmmi5				
\font\fivesy=cmsy5				
\font\fivebf=cmbx5				

\skewchar\fivemi='177
\skewchar\fivesy='60


\font\sixrm=cmr6				
\font\sixi=cmmi6				
\font\sixsy=cmsy6				
\font\sixbf=cmbx6				

\skewchar\sixi='177
\skewchar\sixsy='60


\font\sevenrm=cmr7				
\font\seveni=cmmi7				
\font\sevensy=cmsy7				
\font\sevenit=cmti7				
\font\sevenbf=cmbx7				

\skewchar\seveni='177
\skewchar\sevensy='60


\font\eightrm=cmr8				
\font\eighti=cmmi8				
\font\eightsy=cmsy8				
\font\eightit=cmti8				
\font\eightbf=cmbx8				

\skewchar\eighti='177
\skewchar\eightsy='60


\font\ninei=cmmi9
\font\ninesy=cmsy9

\skewchar\ninei='177
\skewchar\ninesy='60


\font\tenrm=cmr10				
\font\teni=cmmi10				
\font\tensy=cmsy10				
\font\tenex=cmex10				
\font\tenit=cmti10				
\font\tensl=cmsl10				
\font\tenbf=cmbx10				
\font\tentt=cmtt10				
\font\tenss=cmss10				
\font\tensc=cmcsc10				
\font\tenbi=cmmib10				

\skewchar\teni='177
\skewchar\tenbi='177
\skewchar\tensy='60

\def\tenpoint{\ifmmode\err@badsizechange\else
	\textfont0=\tenrm \scriptfont0=\sevenrm \scriptscriptfont0=\fiverm
	\textfont1=\teni  \scriptfont1=\seveni  \scriptscriptfont1=\fivemi
	\textfont2=\tensy \scriptfont2=\sevensy \scriptscriptfont2=\fivesy
	\textfont3=\tenex \scriptfont3=\tenex   \scriptscriptfont3=\tenex
	\textfont4=\tenit \scriptfont4=\sevenit \scriptscriptfont4=\sevenit
	\textfont5=\tensl
	\textfont6=\tenbf \scriptfont6=\sevenbf \scriptscriptfont6=\fivebf
	\textfont7=\tentt
	\textfont8=\tenbi \scriptfont8=\seveni  \scriptscriptfont8=\fivemi
	\def\rm{\tenrm\fam=0 }%
	\def\it{\tenit\fam=4 }%
	\def\sl{\tensl\fam=5 }%
	\def\bf{\tenbf\fam=6 }%
	\def\tt{\tentt\fam=7 }%
	\def\ss{\tenss}%
	\def\sc{\tensc}%
	\def\bmit{\fam=8 }%
	\rm\setparameters\setbaselines\fi}


\font\twelverm=cmr12				
\font\twelvei=cmmi12				
\font\twelvesy=cmsy10	scaled\magstep1		
\font\twelveex=cmex10	scaled\magstep1		
\font\twelveit=cmti12				
\font\twelvesl=cmsl12				
\font\twelvebf=cmbx12				
\font\twelvett=cmtt12				
\font\twelvess=cmss12				
\font\twelvesc=cmcsc10	scaled\magstep1		
\font\twelvebi=cmmib10	scaled\magstep1		

\skewchar\twelvei='177
\skewchar\twelvebi='177
\skewchar\twelvesy='60

\def\twelvepoint{\ifmmode\err@badsizechange\else
	\textfont0=\twelverm \scriptfont0=\eightrm \scriptscriptfont0=\sixrm
	\textfont1=\twelvei  \scriptfont1=\eighti  \scriptscriptfont1=\sixi
	\textfont2=\twelvesy \scriptfont2=\eightsy \scriptscriptfont2=\sixsy
	\textfont3=\twelveex \scriptfont3=\tenex   \scriptscriptfont3=\tenex
	\textfont4=\twelveit \scriptfont4=\eightit \scriptscriptfont4=\sevenit
	\textfont5=\twelvesl
	\textfont6=\twelvebf \scriptfont6=\eightbf \scriptscriptfont6=\sixbf
	\textfont7=\twelvett
	\textfont8=\twelvebi \scriptfont8=\eighti  \scriptscriptfont8=\sixi
	\def\rm{\twelverm\fam=0 }%
	\def\it{\twelveit\fam=4 }%
	\def\sl{\twelvesl\fam=5 }%
	\def\bf{\twelvebf\fam=6 }%
	\def\tt{\twelvett\fam=7 }%
	\def\ss{\twelvess}%
	\def\sc{\twelvesc}%
	\def\bmit{\fam=8 }%
	\rm\setparameters\setbaselines\fi}


\font\fourteenrm=cmr12	scaled\magstep1		
\font\fourteeni=cmmi12	scaled\magstep1		
\font\fourteensy=cmsy10	scaled\magstep2		
\font\fourteenex=cmex10	scaled\magstep2		
\font\fourteenit=cmti12	scaled\magstep1		
\font\fourteensl=cmsl12	scaled\magstep1		
\font\fourteenbf=cmbx12	scaled\magstep1		
\font\fourteentt=cmtt12	scaled\magstep1		
\font\fourteenss=cmss12	scaled\magstep1		
\font\fourteensc=cmcsc10 scaled\magstep2	
\font\fourteenbi=cmmib10 scaled\magstep2	

\skewchar\fourteeni='177
\skewchar\fourteenbi='177
\skewchar\fourteensy='60

\def\fourteenpoint{\ifmmode\err@badsizechange\else
	\textfont0=\fourteenrm \scriptfont0=\tenrm \scriptscriptfont0=\sevenrm
	\textfont1=\fourteeni  \scriptfont1=\teni  \scriptscriptfont1=\seveni
	\textfont2=\fourteensy \scriptfont2=\tensy \scriptscriptfont2=\sevensy
	\textfont3=\fourteenex \scriptfont3=\tenex \scriptscriptfont3=\tenex
	\textfont4=\fourteenit \scriptfont4=\tenit \scriptscriptfont4=\sevenit
	\textfont5=\fourteensl
	\textfont6=\fourteenbf \scriptfont6=\tenbf \scriptscriptfont6=\sevenbf
	\textfont7=\fourteentt
	\textfont8=\fourteenbi \scriptfont8=\tenbi \scriptscriptfont8=\seveni
	\def\rm{\fourteenrm\fam=0 }%
	\def\it{\fourteenit\fam=4 }%
	\def\sl{\fourteensl\fam=5 }%
	\def\bf{\fourteenbf\fam=6 }%
	\def\tt{\fourteentt\fam=7}%
	\def\ss{\fourteenss}%
	\def\sc{\fourteensc}%
	\def\bmit{\fam=8 }%
	\rm\setparameters\setbaselines\fi}


\font\seventeenrm=cmr10 scaled\magstep3		


\newdimen\rp@
\newcount\@basestretchnum
\newskip\@baseskip
\newskip\headskip
\newskip\footskip


\def\setparameters{\rp@=.1em
	\headskip=24\rp@
	\footskip=\headskip
	\delimitershortfall=5\rp@
	\nulldelimiterspace=1.2\rp@
	\scriptspace=0.5\rp@
	\abovedisplayskip=10\rp@ plus3\rp@ minus5\rp@
	\belowdisplayskip=10\rp@ plus3\rp@ minus5\rp@
	\abovedisplayshortskip=5\rp@ plus2\rp@ minus4\rp@
	\belowdisplayshortskip=10\rp@ plus3\rp@ minus5\rp@
	\normallineskip=\rp@
	\lineskip=\normallineskip
	\normallineskiplimit=0pt
	\lineskiplimit=\normallineskiplimit
	\jot=3\rp@
	\setbox0=\hbox{\the\textfont3 B}\p@renwd=\wd0
	\skip\footins=12\rp@ plus3\rp@ minus3\rp@
	\skip\topins=0pt plus0pt minus0pt}


\def\setbaselines{\maxdepth=4\rp@\baselinestretch=\@basestretchnum}


\def\baselinestretch{\afterassignment\@basestretch\@basestretchnum}
\def\@basestretch{%
	\@baseskip=12\rp@ \divide\@baseskip by1000
	\normalbaselineskip=\@basestretchnum\@baseskip
	\baselineskip=\normalbaselineskip
	\bigskipamount=\the\baselineskip
		plus.25\baselineskip minus.25\baselineskip
	\medskipamount=.5\baselineskip
		plus.125\baselineskip minus.125\baselineskip
	\smallskipamount=.25\baselineskip
		plus.0625\baselineskip minus.0625\baselineskip
	\setbox\strutbox=\hbox{\vrule height.708\baselineskip
		depth.292\baselineskip width0pt }}



\def\makeheadline{\vbox to0pt{\baselinestretch=1000
	\vskip-\headskip \vskip1.5pt
	\line{\vbox to\ht\strutbox{}\the\headline}\vss}\nointerlineskip}

\def\makefootline{\baselineskip=\footskip\line{\the\footline}}

\def\big#1{{\hbox{$\left#1\vbox to8.5\rp@ {}\right.\n@space$}}}
\def\Big#1{{\hbox{$\left#1\vbox to11.5\rp@ {}\right.\n@space$}}}
\def\bigg#1{{\hbox{$\left#1\vbox to14.5\rp@ {}\right.\n@space$}}}
\def\Bigg#1{{\hbox{$\left#1\vbox to17.5\rp@ {}\right.\n@space$}}}


\mathchardef\alpha="710B
\mathchardef\beta="710C
\mathchardef\gamma="710D
\mathchardef\delta="710E
\mathchardef\epsilon="710F
\mathchardef\zeta="7110
\mathchardef\eta="7111
\mathchardef\theta="7112
\mathchardef\iota="7113
\mathchardef\kappa="7114
\mathchardef\lambda="7115
\mathchardef\mu="7116
\mathchardef\nu="7117
\mathchardef\xi="7118
\mathchardef\pi="7119
\mathchardef\rho="711A
\mathchardef\sigma="711B
\mathchardef\tau="711C
\mathchardef\upsilon="711D
\mathchardef\phi="711E
\mathchardef\chi="711F
\mathchardef\psi="7120
\mathchardef\omega="7121
\mathchardef\varepsilon="7122
\mathchardef\vartheta="7123
\mathchardef\varpi="7124
\mathchardef\varrho="7125
\mathchardef\varsigma="7126
\mathchardef\varphi="7127
\mathchardef\imath="717B
\mathchardef\jmath="717C
\mathchardef\ell="7160
\mathchardef\wp="717D
\mathchardef\partial="7140
\mathchardef\flat="715B
\mathchardef\natural="715C
\mathchardef\sharp="715D


\def\err@badsizechange{%
	\immediate\write16{--> Size change not allowed in math mode, ignored}}

\baselinestretch=1000
\tenpoint

\catcode`\@=12					
\catcode`\@=11
\expandafter\ifx\csname @iasmacros\endcsname\relax
	\global\let\@iasmacros=\par
\else	\immediate\write16{}
	\immediate\write16{Warning:}
	\immediate\write16{You have tried to input iasmacros more than once.}
	\immediate\write16{}
	\endinput
\fi
\catcode`\@=12


\def\rmb{\seventeenrm}

\def\singlespace{\baselineskip=\normalbaselineskip}
\def\halfspace{\baselineskip=1.5\normalbaselineskip}
\def\doublespace{\baselineskip=2\normalbaselineskip}


\def\AB{\bigskip\parindent=40pt
        \centerline{\bf ABSTRACT}\medskip\halfspace\narrower}
\def\AE{\bigskip\nonarrower\doublespace}
\def\nonarrower{\advance\leftskip by-\parindent
	\advance\rightskip by-\parindent}


\def\boxit#1{\vbox{\hrule\hbox{\vrule\kern3pt
	\vbox{\kern3pt#1\kern3pt}\kern3pt\vrule}\hrule}}

\def\hence{\leavevmode\hbox{\bf .\raise5.5pt\hbox{.}.} }

\def\dalemb#1#2{{\vbox{\hrule height.#2pt
	\hbox{\vrule width.#2pt height#1pt \kern#1pt \vrule width.#2pt}
	\hrule height.#2pt}}}
\def\gtorder{\mathrel{\raise.3ex\hbox{$>$}\mkern-14mu
             \lower0.6ex\hbox{$\sim$}}}
\def\ltorder{\mathrel{\raise.3ex\hbox{$<$}\mkern-14mu
             \lower0.6ex\hbox{$\sim$}}}

\newdimen\fullhsize
\newbox\leftcolumn
\def\twoup{\hoffset=-.5in \voffset=-.25in
  \hsize=4.75in \fullhsize=10in \vsize=6.9in
  \def\fullline{\hbox to\fullhsize}
  \let\lr=L
  \output={\if L\lr
        \global\setbox\leftcolumn=\columnbox\global\let\lr=R \advancepageno
      \else \doubleformat \global\let\lr=L\fi
    \ifnum\outputpenalty>-20000 \else\dosupereject\fi}
  \def\doubleformat{\shipout\vbox{
    \fullline{\box\leftcolumn\hfil\columnbox}\advancepageno}}
  \def\columnbox{\leftline{\vbox{\makeheadline\pagebody\makefootline}}}
  \tolerance=1000 }

\twelvepoint
\doublespace
{\nopagenumbers{
\rightline{IASSNS-HEP-97/12}
\rightline{~~~February 20, 1997}
\bigskip\bigskip
\centerline{\rmb SU(4) Preonic Interpretation of the}
\centerline{\rmb HERA Positron-Jet Events}
\medskip
\centerline{\it Stephen L. Adler
}
\centerline{\bf Institute for Advanced Study}
\centerline{\bf Princeton, NJ 08540}
\medskip
\bigskip\bigskip
\leftline{\it Send correspondence to:}
\medskip
{\singlespace\leftline{Stephen L. Adler}
\leftline{Institute for Advanced Study}
\leftline{Olden Lane, Princeton, NJ 08540}
\leftline{Phone 609-734-8051; FAX 609-924-8399; email adler@sns.ias.
edu}}
\bigskip\bigskip
}}
\vfill\eject
\pageno=2
\AB
We show that our recent $SU(4)$ generalized ``rishon'' composite model
for quarks and leptons, augmented either by the hypothesis of breaking of 
color $SU(3)$ to ``glow'' $SO(3)$, or by the hypothesis of incomplete 
color neutralization in very hard processes, 
leads to an interpretation of the 
HERA positron-jet events as the production by color gluon exchange, 
followed by the decay by color gluon emission, of the positron 
member of a heavy family partner of the positron.  
\AE
\bigskip\bigskip
\vfill\eject
\pageno=3
In a recent paper [1] we have given a composite model of quarks and leptons,  
based on the semisimple gauge group $SU(4)$, with the preons in the 10 
representation.  Our model is a generalization of the ``rishon'' model of 
Harari and Seiberg [2], in which the $\overline{T}_L$  and $\overline{V}_L$ 
rishons correspond respectively   
to the $SU(3)$ representations $6_L$ and $3_L$ appearing 
in the decomposition of the $10_L$ under the symmetry breaking scheme  
$SU(4) \supset SU(3) \times U(1)$.  The $10_L$ also contains a 
color singlet $\overline{S}_L$, which plays no direct role in the binding  
of composite quarks and leptons, but participates in the formuation of a 
postulated condensate $S_L^3T_L^3$, which serves three functions in the model. 
 
The first function of the condensate is to break $SU(4)$ to 
$SU(3) \times U(1)$, providing the basis for the classification of composite 
quarks and leptons in the model according to their $SU(3)$ triality, under   
an assumption that color neutralization to the state of lowest $SU(3)$ 
Casimir always takes place.  
The second function of the condensate is to 
permit the occurrence of the 
weak interactions, through the action of a composite of three gluons 
in the $SU(3)$ representations 3 or $\overline{3}$  contained in the 
15 of $SU(4)$ under $SU(4) \supset SU(3) \times U(1)$, 
accompanied by the action of 
the condensate.  The third function  of the condensate is to break 
the original $Z_{12}$ chiral symmetry of the model to $Z_6$.  As shown 
in [1], the original $Z_{12}$ chiral symmetry leads to the occurrence in 
the model of 6 zero mass quark-lepton families; 
after breaking into $Z_6$, these mix into  three distinct quark-lepton 
families,  two of which remain at zero mass but one of which is no longer 
protected by chiral symmetry from acquiring mass on a scale characterizing 
the condensate.  Thus, the model of [1] corresponds to the experimentally 
observed fact that there are two light families and one heavy family in the 
standard model, and {\it makes the prediction  
that for each of the three standard model families, there is a heavy partner 
occurring at the scale of the condensate.}  In particular, the positron $e^+$ 
would have a heavy family partner $E^+$ at a mass scale characterizing the  
condensate, which we assume in the further discussion to be of order 
the top quark mass.  

The occurrence of heavy family partners of the light families cannot in 
itself account for the excess positron-jet events recently reported [3] by 
the H1 and ZEUS collaborations at HERA.  The reason is that 
if color $SU(3)$ remains 
an exact symmetry,  and if color neutralization is assumed to be 
instantaneous, 
the $E^+$ would be produced as a color singlet, and its 
dominant decay mode would be the  
electromagnetic decay $E^+ \to e^+ + \gamma$, which does not corresond to  the 

$e^+ + {\rm jet}$ signature reported by the HERA groups.  However, the 
production and decay modes of the $E^+$ in the model of [1] depend strongly 
on the details of the surviving unbroken symmetries, and of color   
neutralization.  

Suppose, for example, 
that color $SU(3)$ were weakly broken to ``glow'' $SO(3)$, while 
maintaining triality conservation modulo 3, as suggested by Slanksy, 
Goldman, and Shaw [4], with the threshold for excitation of free color 
of order the top quark mass.  [It makes sense to talk about a free color 
threshold in this context because under $SU(3) \supset SO(3)$, 
all states of $SU(3)$ decompose 
into diality zero integer $L$ representations of $SO(3)$; 
half-integer, nonzero diality 
representations of the covering group $SU(2)$, which are confined, are never 
encountered.]
The possibility of the breaking of color to glow can be naturally 
incorporated into the model of [1] for two reasons.  First of all, the 
state classification of [1] uses only the conservation of triality modulo 3, 
and not the details of the particular $SU(3)$ representations corresponding 
to each triality sector, so a triality preserving breaking of $SU(3)$ 
to $SO(3)$ leaves this classification, and in particular the distinction 
between leptons and quarks, intact.  Second, to achieve a triality preserving 
breaking of $SU(3)$ to $SO(3)$, the smallest $SU(3)$ representation 
with nonzero vacuum expectation must be [4] the 27, since this is the 
smallest triality zero representation of $SU(3)$ that contains an $SO(3)$ 
singlet.  But since the postulated condensate $S_L^3T_L^3$ of our model 
has the $SU(3)$ tensor product structure $6 \times 6 \times 6$, 
which contains the 27, it is consistent with the framework of [1] to 
additionally postulate that the condensate has a small component in the 
27 representation of color $SU(3)$ as well as a dominant color singlet 
component. 
The effect of  a breaking of color $SU(3)$ to glow $SO(3)$, with a free  
color excitation threshold of order the top quark mass, would be to leave   
the standard model leptons as very nearly color singlets (particularly 
so for the electron, because of its very small mass), 
but to permit their heavy counterparts to carry significant admixtures 
of color non-singlet states.  In this scenario, the $E^+$ could carry a 
color octet component, and the dominant decay mode would then be single 
color gluon emission $E^+ \to e^+ + g$, which would appear as a positron 
plus gluon jet final state, corresponding to the signature observed at 
HERA.  The postulated breaking of color to glow is also relevant for heavy 
family production processes, because if the $E^+$ carries a color octet 
component, the reaction $e^+ + u/d \to E^+ + u/d$ can proceed by single 
color gluon exchange, leading to a larger production cross section than 
would be expected for production by photon or $Z^0$ exchange.  

An alternative scenario, which does not require color $SU(3)$ to be 
broken, is simply to assume that color neutralization does not fully 
take place in the hard processes involved in $E^+$ production and its 
subsequent decay.  Recall that in the model of [1], the $e^+$ and 
$E^+$ are both $TTT$ three preon bound states, with the $SU(3)$ 
wave function (before color neutralization) corresponding to $6_L \times
6_L \times 6_R$, which has the Clebsch series $8 + 10 + \overline {10}+...$ 
and contains no color singlet.  The postulate made in [1] is that color 
neutralization occurs by picking up color gluons from the vacuum until the 
$SU(3)$ state with lowest Casimir is attained, so that only 
the $SU(3)$ triality plays a role in enumerating the possible states.   
However, in very hard processes, characterized by momentum transfers
much larger than the QCD scale, it is possible that this color neutralization 
could be incomplete, and that the $E^+$ would then behave as a state 
with the color wave function suggested by the preon model Clebsch series.  
Again, as in the color breaking scenario, 
this would permit the production of the $E^+$ from positrons by 
single color gluon exchange, and its subsequent rapid decay into a positron 
and a gluon jet.  

In summary, we suggest that  
the production and decay of the excess HERA events can be accounted 
for by the composite model of [1], augmented by either the assumption that the 

$Z_6$ condensate that breaks $SU(4)$ to color $SU(3)$ contains 
a small component that further breaks color $SU(3)$ to glow $SO(3)$, or by 
the assumption that color symmetry remains exact but that 
the color neutralization postulated in [1] is incomplete 
in hard processes.

\bigskip
\centerline{\bf Acknowledgments}
I wish to thank Frank Wilczek for forwarding an email describing the 
HERA seminar of 12/19/97, and members of the high energy physics 
group at the IAS for stimulating discussions.  
This work was supported in part by the Department of Energy under
Grant \#DE--FG02--90ER40542.
\vfill\eject
\centerline{\bf References}
\bigskip
\noindent
\item{[1]}  S. L. Adler, ``Frustrated SU(4) as the Preonic Precursor 
of the Standard Model'', IASSNS-HEP-96/104, submitted to Nuclear Physics. 
Available on the bulletin board as hep-th/9610190; original version 
posted 10/25/96; revised version posted 2/20/97.\hfill\break  
\bigskip 
\noindent
\item{[2]}  H. Harari and N. Seiberg, Phys. Lett. 98B (1982) 269; 
Nucl. Phys. B204 (1982) 141; Phys. Lett. 102B (1981) 263.\hfill\break  
\bigskip
\noindent
\item{[3]}  H1 and ZEUS collaborations at HERA; data presented at a 
seminar on 2/19/97, submitted for publication to Zeitschrift f\"ur Physik.  
\hfill\break
\bigskip
\noindent
\item{[4]} R. Slansky, T. Goldman, and G. L. Shaw, Phys. Rev. Lett. 47  
(1981) 887. \hfill\break
\bigskip
\bigskip
\bye